\documentclass[12pt,preprint]{aastex}
\begin{document}

\def\wisk#1{\ifmmode{#1}\else{$#1$}\fi}

\def\lt     {\wisk{<}}
\def\gt     {\wisk{>}}
\def\le     {\wisk{_<\atop^=}}
\def\ge     {\wisk{_>\atop^=}}
\def\lsim   {\wisk{_<\atop^{\sim}}}
\def\gsim   {\wisk{_>\atop^{\sim}}}
\def\kms    {\wisk{{\rm ~km~s^{-1}}}}
\def\Lsun   {\wisk{{\rm L_\odot}}}
\def\Zsun   {\wisk{{\rm Z_\odot}}}
\def\Msun   {\wisk{{\rm M_\odot}}}
\def\um     {$\mu$m}
\def\mic     {\mu{\rm m}}
\def\sig    {\wisk{\sigma}}
\def\etal   {{\sl et~al.\ }}
\def\eg     {{\it e.g.\ }}
 \def\ie     {{\it i.e.\ }}
\def\bsl    {\wisk{\backslash}}
\def\by     {\wisk{\times}}
\def\half {\wisk{\frac{1}{2}}}
\def\third {\wisk{\frac{1}{3}}}
\def\nwm2sr {\wisk{\rm nW/m^2/sr\ }}
\def\nw2m4sr {\wisk{\rm nW^2/m^4/sr\ }}

\title{A measurement of large-scale peculiar velocities of clusters of galaxies: results and cosmological implications.}

\author{A. Kashlinsky\altaffilmark{1}, F.  Atrio-Barandela\altaffilmark{2},
  D. Kocevski\altaffilmark{3}, H.  Ebeling\altaffilmark{4}}
\altaffiltext{1}{SSAI and Observational Cosmology Laboratory, Code 665, Goddard
  Space Flight Center, Greenbelt MD 20771; e--mail:
  alexander.kashlinsky@nasa.gov} \altaffiltext{2}{Fisica Teorica,
  University of Salamanca, 37008 Salamanca, Spain} \altaffiltext{3}{Department
  of Physics, University of California at Davis, 1 Shields Avenue, Davis, CA
  95616} \altaffiltext{4}{Institute for Astronomy, University of Hawaii, 2680
  Woodlawn Drive, Honolulu, HI 96822 }

\begin{abstract}
{Peculiar velocities of clusters of galaxies can be measured by
studying the fluctuations in the cosmic microwave background (CMB)
generated by the scattering of the microwave photons by the hot
X-ray emitting gas inside clusters. While for individual clusters
such measurements result in large errors, a large statistical
sample of clusters allows one to study cumulative quantities
dominated by the overall bulk flow of the sample with the
statistical errors integrating down. We present results from such
a measurement using the largest all-sky X-ray cluster catalog
combined to date and the 3-year WMAP CMB data. We find a strong
and coherent bulk flow on scales out to at least $\gsim 300
h^{-1}$Mpc, the limit of our catalog. This flow is difficult to
explain by gravitational evolution within the framework of the
concordance $\Lambda$CDM model and may be indicative of the tilt
exerted across the entire current horizon by far-away
pre-inflationary inhomogeneities.}
\end{abstract}

In the gravitational-instability picture peculiar velocities probe
directly the peculiar gravitational potential [e.g. Kashlinsky \&
Jones 1991, Strauss \& Willick 1995]. Inflation-based theories,
such as the concordance $\Lambda$CDM model, predict that, on
scales outside the horizon during the radiation-dominated era, the
peculiar density field remained in the Harrison-Zeldovich regime
set during inflationary epoch and on these scales, the peculiar
bulk velocity due to gravitational instability should decrease as
$V_{\rm rms} \propto r^{-1}$ and be quite small. Peculiar
velocities can be obtained from the kinematic SZ (KSZ) effect on
the CMB photons by the hot gas in clusters of galaxies [e.g.
Birkinshaw 1999]. For each cluster the KSZ term is small, but
measuring a quantity derived from CMB data for a sizeable ensemble
of many clusters moving at a coherent bulk flow can, however,
overcome this limitation. As proposed by Kashlinsky \&
Atrio-Barandela (2000, KA-B), such a measurement will be dominated
by the bulk-flow KSZ component with other contributions
integrating down. This quantity, {\it the dipole of the cumulative
CMB temperature field evaluated at cluster positions}, is used in
this investigation of the 3-year WMAP data together with the
largest X-ray selected sample of clusters to date to obtain the
best measurement yet of bulk flows out to scales of $\gsim 300
h^{-1}$Mpc. Technical details of the analysis are given in the
companion paper (Kashlinsky et al 2008 - KA-BKE). Our findings
imply that the Universe has a surprisingly coherent bulk motion
out to at least $\simeq 300h^{-1}$Mpc and with a fairly high
amplitude of $\gsim$600-1000 km/sec, necessary to produce the
measured amplitude of the dipole signal of $\simeq$2-3$\mu$K. Such
a motion is difficult to account for by gravitational instability
within the framework of the standard concordance $\Lambda$CDM
cosmology but could be explained by the gravitational pull of
pre-inflationary remnants located well outside the present-day
horizon.

\section{Method and data preparation}

If a cluster at angular position $\vec{y}$ has the line-of-sight
velocity $v$ with respect to the CMB, the CMB fluctuation caused
by the SZ effect at frequency $\nu$ at this position will be
$\delta_\nu(\vec y)=\delta_{\rm TSZ}(\vec y)G(\nu)+ \delta_{\rm
KSZ}(\vec y)H(\nu)$, with $ \delta_{\rm TSZ}$=$\tau T_{\rm
X}/T_{\rm e,ann}$ and $\delta_{\rm KSZ}$=$\tau v/c$. Here
$G(\nu)\simeq-1.85$ to $-1.25$ and $H(\nu)\simeq 1$ over the WMAP
frequencies, $\tau$ is the projected optical depth due to Compton
scattering, $T_{\rm X}$ is the temperature of the intra-cluster
gas, and $k_{\rm B}T_{\rm e,ann}$=511 keV. Averaged over many
isotropically distributed clusters moving at a significant bulk
velocity with respect to the CMB, the dipole from the kinematic
term will dominate, allowing a measurement of $V_{\rm bulk}$. Thus
KA-B suggested measuring the dipole component of $\delta_\nu(\vec
y)$.

We use a normalized notation for the dipole power $C_{1}$, such
that a coherent motion at velocity $V_{\rm bulk}$ leads to
$C_{1,{\rm kin}}= T_{\rm CMB}^2 \langle \tau \rangle^2 V_{\rm
bulk}^2/c^2$, where $T_{\rm CMB} =2.725$K. For reference,
$\sqrt{C_{1,{\rm kin}}}\simeq 1 (\langle \tau \rangle/10^{-3})
(V_{\rm bulk}/100{\rm km/sec}) \; \mu$K. When computed from the
total of $N_{\rm cl}$ positions, the dipole will also have
positive contributions from 1) instrument noise, 2) the thermal SZ
(TSZ) component, 3) the cosmological CMB fluctuation component
arising from the last-scattering surface, and 4) the various
foreground components within the WMAP frequency range. The last of
these contributions can be significant at the lowest WMAP
frequencies (channels K \& Ka) and, hence, we restrict this
analysis to the WMAP Channels Q, V \& W which have negligible
foreground contributions. The contributions to the dipole from the
above terms can be estimated as
$\langle\delta_\nu(\vec{y})\cos\theta\rangle$ at the $N_{\rm cl}$
different cluster locations with polar angle $\theta$. For $N_{\rm
cl}\gg1$ the dipole of $\delta_\nu$ becomes $a_{1m} \simeq
a_{1m}^{\rm kin} +a_{1m}^{\rm TSZ} + a_{1m}^{\rm CMB} +
\frac{\sigma_{\rm noise}}{\sqrt{N_{\rm cl}}}$. Here $a_{1m}^{\rm
CMB}$ is the residual dipole produced at the cluster locations by
the primordial CMB anisotropies. The dipole power is $C_1=
\sum_{m=-1}^{m=1} |a_{1m}|^2$. The notation for $a_{1m}$ is such
that $m$=$0,1,-1$ correspond to the $(x,y,z)$ components, with $z$
running perpendicular to the Galactic plane towards the NGP, and
$(x,y)$ being the Galactic plane with the $x$-axis passing through
the Galactic center. This dipole signal should not be confused
with the "global CMB dipole" that arises from our {\it local}
motion relative to the CMB. The kinematic signal investigated here
does not contribute significantly to the ``global CMB dipole"
arising from only a small number of pixels. When the latter is
subtracted from the original CMB maps, only a small fraction,
$\sim (N_{\rm cl}/N_{\rm total})\lsim 10^{-3}$, of the kinematic
signal $C_{1,{\rm kin}}$ is removed.

The TSZ dipole for a random cluster distribution is $a_{1m}^{\rm
TSZ}\sim(\langle \tau T_{\rm  X}\rangle/T_{\rm e,ann}) N_{\rm
cl}^{-1/2}$ decreasing with increasing $N_{\rm cl}$. This decrease
could be altered if clusters are not distributed randomly and
there may be some cross-talk between the monopole and dipole terms
especially for small/sparse samples \cite{watkins-feldman}, but
the value of the TSZ dipole will be estimated directly from the
maps as discussed below and in greater detail in Kashlinsky et al
(2008 - KA-BKE). The residual CMB dipole, $C_{1,{\rm CMB}}$, will
exceed $\sigma_{\rm CMB}^2/N_{\rm cl}$ because the intrinsic
cosmological CMB anisotropies are correlated. On the smallest
angular scales in the WMAP data $\sigma_{\rm CMB}\simeq 80\mu$K,
so these anisotropies could be seen as the largest dipole noise
source. However, because the power spectrum of the underlying CMB
anisotropies is accurately known, this component can be removed
with a filter described next.

To remove the cosmological CMB anisotropies we filtered each
channel maps separately with the Wiener filter as follows. With
the known power spectrum of the cosmological CMB fluctuations,
$C_\ell^{\Lambda{\rm CDM}}$, a filter $F_\ell$ in $\ell$-space
which minimizes $\langle (\delta T - \delta_{\rm instrument \;
noise})^2\rangle$ in the presence of instrument noise is given by
$F_\ell = (C_\ell - C_\ell^{\Lambda{\rm CDM}})/C_\ell$, with
$C_\ell$ being the measured power spectrum of each map. Convolving
the maps with $F_\ell$ minimizes the contribution of the
cosmological CMB to the dipole. The maps for {\it each} of the
eight WMAP channels were thus processed as follows: 1) for
$C_\ell^{\Lambda {\rm CDM}}$ we adopted the best-fit cosmological
model for the WMAP data \cite{hinshaw} available from
http://lambda.gsfc.nasa.gov; 2) each map was decomposed into
multipoles, $a_{\ell m}$, using HEALPix \cite{healpix}; 3) the
power spectrum of each map, $C_\ell$, was then computed and
$F_\ell$ constructed; 4) the $a_{\ell m}$ maps were multiplied by
$F_\ell$ and Fourier-transformed back into the angular space
$(\theta,\phi)$. We then removed the intrinsic dipole, quadrupole
and octupole. The filtering affects the effective value of $\tau$
for each cluster and we calculate this amount later.

Here we use an all-sky cluster sample created by combining the
ROSAT-ESO Flux Limited X-ray catalog (REFLEX) \cite{bohringer} in
the southern hemisphere, the extended Brightest Cluster Sample
(eBCS) \cite{ebeling1,ebeling2} in the north, and the Clusters in
the Zone of Avoidance (CIZA) \cite{ebeling3,kocevski1} sample
along the Galactic plane. These are the most statistically
complete X-ray selected cluster catalogs ever compiled in their
respective regions of the sky. All three surveys are X-ray
selected and X-ray flux limited using RASS data. The creation of
the combined all-sky catalogue of 782 clusters is described in
detail by \cite{kocevski2} and KA-BKE.

We started with 3-year ``foreground-cleaned" WMAP data
(http://lambda.gsfc.nasa.gov) in each differencing assembly (DA)
of the Q, V, and W bands. Each DA is analyzed separately giving us
eight independent maps to process: Q1, Q2, V1, V2, W1,..., W4. The
CMB maps are pixelized with the HEALPix parameter $N_{\rm
side}$=512 corresponding to pixels $\simeq 7^\prime$ on the side
or pixel area $4\times 10^{-6}$ sr. This resolution is much
coarser than that of the X-ray data, which makes our analysis
below insensitive to the specifics of the spatial distribution of
the cluster gas, such as cooling flows, deviations from spherical
symmetry, etc. In the filtered maps for each DA we select all WMAP
pixels within the total area defined by the cluster X-ray
emission, repeating this exercise for cluster subsamples
populating cumulative redshift bins up to a fixed $z$. In order to
eliminate the influence of Galactic emission and non-CMB radio
sources, the CMB maps are subjected to standard WMAP masking. The
results for the different masks are similar and agree well within
their statistical uncertainties.

The SZ effect ($\propto n_e$, the electron density) has larger
extent than probed by X-rays (X-ray luminosity $\propto n_e^2$),
which is confirmed by our TSZ study using the same cluster
catalogue (AKKE). As shown in AKKE, KA-BKE contributions to the
TSZ signal are detected out to $\gsim 30^\prime$. What is
important in the present context, is that the X-ray emitting gas
is distributed as expected from the $\Lambda$CDM profile
\cite{nfw} scaling as $n_e\propto r^{-3}$ in outer parts. In order
to be in hydrostatic equilibrium such gas must have temperature
decreasing with radius \cite{komatsu}. Indeed, the typical
polytropic index for such gas would be $\gamma \simeq$1.2, leading
to the X-ray temperature decreasing as $T_{\rm X} \propto
n_e^{\gamma-1} \propto r^{-0.6}$ at outer radii. This $T_{\rm X}$
decrease agrees with simulations of cluster formation within the
$\Lambda$CDM model \cite{simulations_temp} and with the available
data on the X-ray temperature profile \cite{pratt}. For such gas,
the TSZ monopole ($\propto T_{\rm X} \tau$) decreases faster than
the KSZ component ($\propto \tau$) when averaged over a
progressively increasing cluster area. To account for this, we
compute the dipole component of the final maps for a range of
effective cluster sizes, namely $[1,2,4,6] \theta_{\rm X-ray}$ and
then the maximal cluster extent is set at $30^\prime$ to avoid a
few large clusters (eg. Coma) bias the dipole determination. We
note that at the final extent our clusters effectively have the
same angular radius of 0.5$^\circ$. Our choice of the maximal
extent is determined by the fact that the SZ signal is detectable
in our sample out to that scale (AKKE), which is $\sim$(3-4)Mpc at
the mean redshift of the sample. Increasing the cluster radius
further to 1$^\circ-3^\circ$, causes the dipole to start
decreasing with the increasing radius, as expected if the pixels
outside the clusters are included diluting the KSZ signal
(KA-BKE).

To estimate uncertainties in the signal only from the clusters, we
use the rest of the map for the distribution and variance of the
noise in the measured signal. We use two methods to preserve the
geometry defined by the mask and the cluster distribution: 1)
$N_{\rm cl}$ central random pixels are selected outside the mask
away from the cluster pixels adding pixels within each cluster's
angular extent around these central random pixels, iteratively
verifying that the selected areas do not fall within either the
mask or any of the known clusters. 2) We also use a slightly
modified version of the above procedure in order to test the
effects of the anisotropy of the cluster catalog. There the
cluster catalog is rotated randomly, ensuring that the overall
geometry of the cluster catalog is accurately preserved. Both
methods yield very similar uncertainties; for brevity, we present
results obtained with the first method.

\section{Results}

Fig. \ref{fig:c1} summarizes our results averaged over all eight
DA's. We find a statistically significant dipole component
produced by the cluster pixels for the spheres and shells
extending beyond $z\simeq 0.05$. It persists {\it as the monopole
component vanishes} and its statistical significance gets
particularly high for the $y$-component. The signal appears only
at the cluster positions and, hence, cannot originate from
instrument noise, the CMB or the remaining Galactic foreground
components, the contributions from which are given by the
uncertainties evaluated from the rest of the CMB map pixels. The
signal is restricted to the cluster pixels and thus must arise
from the two components of the SZ effect, thermal and/or
kinematic.

The TSZ component, however, is given by the monopole term at the
cluster positions and cannot be responsible for the detected
signal. For the largest apertures it vanishes within the small,
compared to the measured dipole, statistical uncertainty, and yet
the dipole term remains large and statistically significant. This
is the opposite of what one should expect if the dipole is
produced by the TSZ component. Any random distribution, such as
TSZ emissions, would generate dipole $\propto \langle \tau T_{\rm
X}\cos \theta\rangle$ which can never exceed (and must be much
less than) the monopole component of that distribution, $\propto
\langle \tau T_{\rm X} \rangle$. On the other hand, any coherent
bulk flow would produce dipole $\propto V_{\rm bulk}\langle \tau
\cos^2 \theta\rangle$, which is bounded from below by the
amplitude of the motion. Furthermore, we find significant dipole
from (at least) $z_{\rm mean}$=0.035 (135 clusters) all the way to
$z_{\rm mean}$=0.11 (674 clusters); its parameters do not depend
on the numbers of clusters, pixels used etc. Any dipole component
arising from the TSZ term would depend on these parameters as it
reflects the (random) dipole of the cluster sample and should thus
decrease as more clusters are added in spheres out to
progressively larger $z$. To verify this, we compute the expected
monopole and dipole terms produced by the TSZ effect using the
parameters of our cluster catalog as discussed in KA-BKE and
recover the monopole term fairly accurately when the
$\beta$-profile assumption is reasonable. The TSZ dipole component
is then a small fraction of the monopole term. When normalized to
the remaining monopole term in Fig.~\ref{fig:c1}a it is completely
negligible compared to the measured dipole. Further, the TSZ
dipole becomes progressively more negligible as more clusters are
added in at higher $z$, and its direction varies randomly
reflecting the random nature of the intrinsic cluster sample
dipole on these large scales. All this is contrary to what we
measure.

We thus conclude that the dipole originates from the KSZ effect
due to the bulk flow of the cluster sample. Our results indicate a
statistically significant bulk-flow component in the final
filtered maps for cluster samples in the $z$-bins from $z\leq$0.05
to $\leq$0.3 corresponding to median depth to $z\simeq$0.1, and it
also persists when the dipole in shells is computed selecting only
clusters at $z\geq$0.12 (the median redshift for this sub-sample
is $\simeq$0.18). Fig.~\ref{fig:c1}e shows that the bulk flow
results in a CMB dipole with little variation - within the
statistical uncertainties - between $z_{\rm median}\simeq 0.03$
and $\gsim$0.12. The monopole component reflects the residual TSZ
contribution which is very small for the maximal cluster aperture
as Fig.~\ref{fig:c1}a shows. (At lower $z$ there may still be some
residual TSZ component, which would be consistent with the more
nearby clusters having a larger {\it angular} SZ extent than the
more distant ones).

To translate the CMB dipole in $\mu$K into $V_{\rm bulk}$ in
km/sec, we generated CMB temperatures produced by the KSZ effect
for each cluster and estimate the dipole amplitude, $C_{1,100}$,
contributed by each 100 km/sec of bulk-flow (KA-BKE). The results
are shown in the last column of Table 1 of KA-BKE for the central
values of the direction of the measured flow; within the
uncertainties of $(l,b)$ they change by at most a few percent. A
bulk flow of 100 km/sec leads to $\sqrt{C_{1,100}}\simeq 1\mu$K
for unfiltered clusters assuming the $\beta$-model; this
corresponds to an average optical depth of our cluster sample of
$\langle \tau \rangle \simeq 10^{-3}$ expected for a typical
galaxy cluster. For NFW clusters the value of $C_{1,100}$ would be
{\it smaller}. Filtering reduces the effective $\tau$ by a factor
of $\simeq 3$. Since a $\beta$-model provides a poor fit to the
measured TSZ component outside the estimated values of
$\theta_{\rm X-ray}$, we compute $C_{1,100}$ within that aperture
where the central value of the bulk-flow dipole has approximately
the same value as at the final apertures. Due to the large size of
our cluster sample ($N_{\rm cl} \sim$130-675), the random
uncertainties in the estimated values of $C_{1,100}$ should be
small, but we cannot exclude a systematic offset related to
selection biases affecting our cluster catalog at high $z$. Such
offset, if present, will become quantifiable with the next version
of our X-ray cluster catalog (in preparation) using the
empirically established SZ profile rather than the current
$\beta$-model to parameterize the cluster TSZ. The good agreement
between the various TSZ-related quantities shown in KA-BKE for
$\theta_{\rm SZ}$=$\theta_{\rm X-ray}$ and the observed values for
both unfiltered and filtered maps suggests, however, that these
systematic uncertainties are likely to be small. They only affect
the accuracy of the determination of the amplitude of the bulk
flow, but not its existence established by the CMB dipole at the
cluster locations. Since the filtering removes $\tau$ in the
outskirts of clusters more effectively, a larger amount of power
is removed in the $\beta$-model when the cluster SZ extent is
increased beyond $\theta_{\rm X-ray}$, than in the steeper profile
measured by AKKE. Thus the effective $\tau$ is possibly
underestimated by using a $\beta$-model, but it cannot exceed (and
must be much less than) the calibration obtained from the
unfiltered $\beta$-model.

\section{Cosmological implications}

\begin{figure}
\plotone{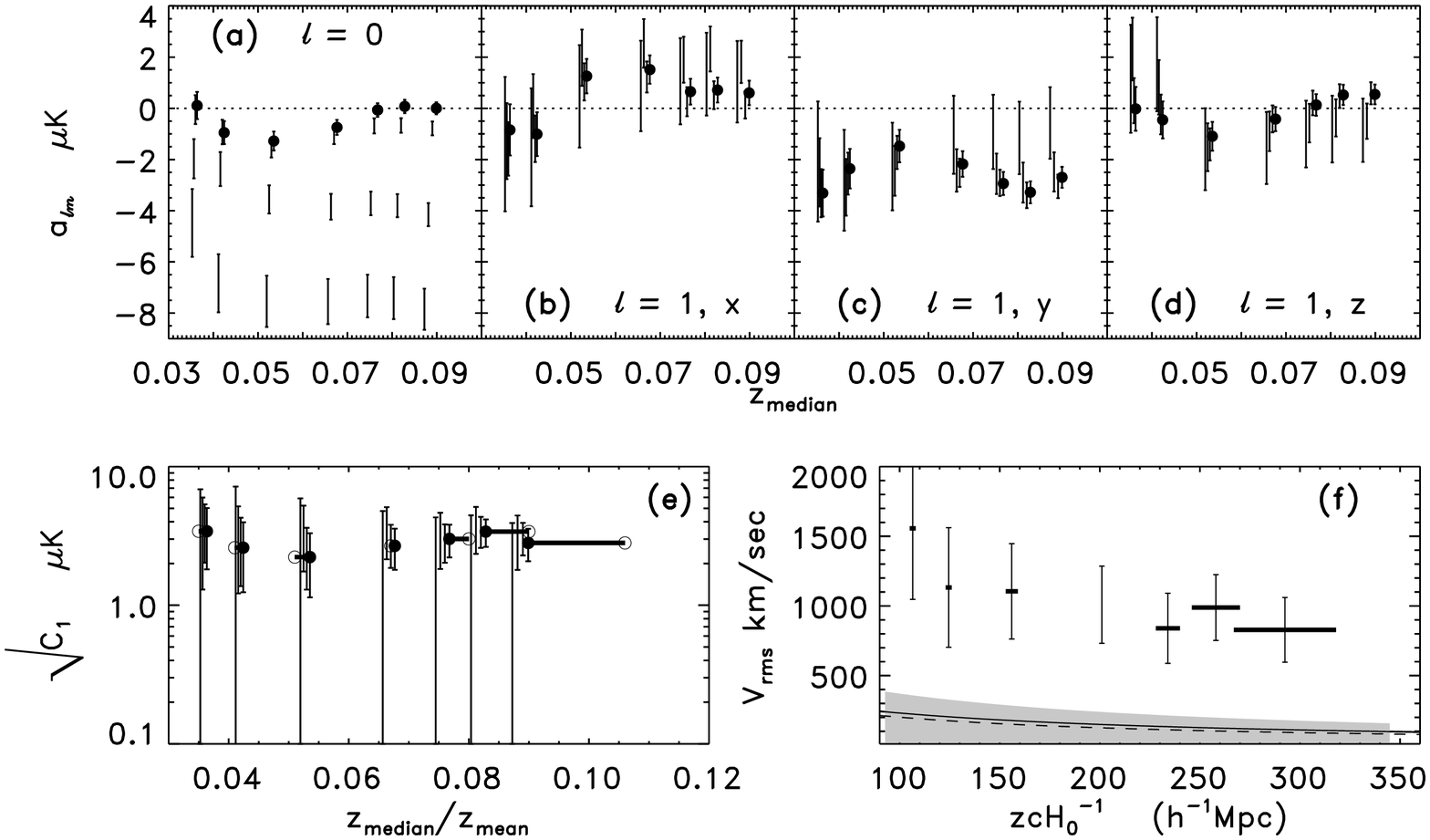} \caption{{\bf Upper panels}: (a)-(d) Monopole and
dipole terms for $\theta_{\rm SZ}=\min[(1,2,4, 6)\times
\theta_{\rm X-ray},30^\prime]$ with 1-$\sigma$ standard
deviations. Values at the maximal aperture, which have the lowest
monopole term, are marked with filled circles. The averaged
monopole and dipole components are weighted with statistical
uncertainties. The statistical significance of the KSZ component
improves as more of the cluster pixels producing the signal are
included at higher $z$. The noise of our measurement of the dipole
at $1.8 (N_{\rm cl}/100)^{-1/2}\mu$K with three-year WMAP data is
in good agreement with the expectations of KA-B. {\bf Lower
panel}. (e) - Outer $z$-bins with signal measured at $\gsim
2\sigma$. Filled circles show the values from Table 1 of KA-BKE at
the maximal aperture vs the median $z$; open symbols show the same
vs the mean $z$. The two symbols are connected to show the
uncertainty in the scale on which the flow is probed. Signal
recovered at $\theta_{\rm SZ} = \min[(1,2,4) \times \theta_{\rm
X-ray},30^\prime]$ is shown with $1\sigma$ error bars; from left
to right in order of increasing aperture. The values are slightly
displaced around the true $z_{\rm median}$ for clearer display.
(f) - Comparison between theoretically expected bulk flow and the
measurements. The rms bulk velocity for the concordance
$\Lambda$CDM model which best fits the WMAP 3-year data for
top-hat (solid line) and Gaussian (dashes) windows; shaded region
marks the 95\% cl from cosmic variance. The results of this study,
translated into km/sec using $\sqrt{C_{1,100}}=0.3\mu$K, are shown
with 1-$\sigma$ errors vs the mean/median redshift of the clusters
in each cumulative $z$-bin. The horizontal bars connect $z_{\rm
mean}$ with $z_{\rm median}$. The results in shells from Table 1
in KA-BKE are omitted in this comparison because of the
theoretical windows plotted, but they show that the motion extends
to mean redshift $\gsim 0.18$ well beyond the horizontal range of
the figure. } \label{fig:c1}
\end{figure}

Conventionally, the entire peculiar velocity field is assumed to
be driven by the peculiar gravitational potential. For a given
cosmological model, the details of the velocity field also depend
on the window function of the dataset. Constructing the precise
window function is beyond the scope of this paper, but the overall
conclusions would be insensitive to its details because the
amplitude and coherence length of the measured flows are quite
unexpected within the concordance $\Lambda$CDM model.
Fig.~\ref{fig:c1}f shows the rms prediction, $\sigma_V$, of the
concordance $\Lambda$CDM model. If produced by gravitational
instability within the concordance $\Lambda$CDM model, the motion
would require the local Universe out to $\sim 300 h^{-1}$Mpc to be
atypical at the level of many standard deviations of the model.
Indeed a value of $\sqrt{C_{1,100}}\sim 3\mu$K is required to
reach peculiar velocities of order 100 km/sec on the relevant
scales. This is much greater than $C_{1,100}$ deduced from the
{\it unfiltered} X-ray data and even then it would be difficult to
explain the approximate constancy of the measured dipole with
depth.

Cosmic variance does not change these conclusions significantly.
For a Gaussian density field the peculiar velocity distribution on
linear scales is Maxwellian, with the probability density of
measuring a 1-D bulk velocity $p(V) dV\! \propto\!
V^2\exp(-1.5V^2/\sigma_V^2)dV$. The probability of finding a
region with $V\!<\!V_0$ is then
$P(V_0)$=$\Gamma(\frac{3}{2},\frac{3V_0^2}{2\sigma_V^2})$ where
$\Gamma$ is the incomplete gamma-function normalized to
$\Gamma(n,\infty)$=1. The (68\%, 95\%) c.l. require
$V_0$=(1.08,1.6)$\sigma_V$; the shaded area in Fig. \ref{fig:c1}f
shows the 95\% c.l. region. In the concordance $\Lambda$CDM model
$\sigma_V$=(150,109) km/sec at $(200,300)h^{-1}$Mpc, so 95\% of
cosmic observers should measure bulk flow velocities less than
(240,180) km/sec at these scales. To make these  numbers
consistent with our measurements - at these scales alone - would
require $\sqrt{C_{1,100}}\gsim 2\mu$K. This is much higher than
even the calibration values for unfiltered data (computed for the
$\beta$-model which leads to a {\it larger} $C_{1,100}$ value than
the NFW-profile clusters) and cannot be accounted for by any
systematic uncertainties in our calibration procedure. Indeed, the
distribution of errors is approximately Gaussian (see Fig. 5 of
KA-BKE), so the probability of measuring velocity
$V_i\pm\epsilon_i$ at scale $r_i$ is ${\cal P}_i \propto
\int_0^\infty P(V) \exp[-\frac{(V-V_i)^2}{2\epsilon_i^2}] dV$
(e.g. Gorski 1991). The probability of several such independent
measurements is the product of ${\cal P}_i$'s; for the numbers
plotted in Fig. 1f this probability is completely negligible. (The
measurements in Fig. 1f are not strictly independent since each
subsequent $z$-bin also contains the clusters from the previous
bin; nevertheless each ${\cal P}_i$ is so small that the overall
probability is still negligible).

The coherence length of the measured bulk flow shows no signs of
convergence out to $\gsim 300 h^{-1}$Mpc, and it is quite possible
that it extends to much larger scales, possibly all the way across
our horizon. An interesting, if exotic, explanation for such a
``dark flow" would come naturally within certain inflationary
models. In general, within these models the observable Universe
represents part of a homogeneous inflated region embedded in an
inhomogeneous space-time. On scales much larger than the Hubble
radius, pre-inflationary remnants can induce tilt including CMB
anisotropies generated by the Grischuk-Zeldovich \cite{gz} effect
\cite{turner,ktf}. These can arise from the parts of space-time
that inflated at different times and rates and would manifest
themselves mainly in the quadrupole component, $Q$: an
inhomogeneity of amplitude $\delta_L\sim 1$ at a distance $L\gg
cH_0^{-1}$ generates a quadrupole $Q\sim \delta_L
(cH_0^{-1}/L)^2$. Consistency with the observed low value of $Q$
would require a sufficiently large number of the inflation's
e-foldings, making the Universe flat to within $|1-\Omega_{\rm
total}| \leq Q$ and causing the scale of inhomogeneity $L$ to
become comparable to the curvature radius ($> 500 cH_0^{-1}$)
\cite{ktf}. Such a tilted universe would lead to a uniform flow
across the observed horizon due to the density gradient produced
by this superhorizon mode. The bulk motion would have an amplitude
of $v \sim c \delta_L (cH_0^{-1}/L)$ and would not generate a
primordial dipole CMB component \cite{turner}. Since the
quadrupole produced by such a pre-inflationary remnant is $Q \sim
(v/c) (cH_0^{-1}/L)$, it is possible for such inhomogeneities to
generate the required motions and be consistent with the observed
value of $Q$ and flatness. Although it would require accidental
alignment, the contribution from such an inhomogeneity to CMB
anisotropies via the GZ effect might, interestingly, also explain
the observed low value of the CMB quadrupole (and possibly also
octupole) compared to the concordance $\Lambda$CDM model.  This
explanation for the measured bulk flow would, however, still
require peculiar velocities generated by gravitational instability
acting on the $\Lambda$CDM density field, which would provide a
random component around the uniform bulk flow. On sufficiently
large scales, such a flow would be ``cold" in the sense that it
would be characterized by a large Mach number \cite{mach1}, which
may be measurable in future cluster surveys \cite{mach2}; the Mach
number then should increase linearly with scale on scales $\gsim
100 h^{-1}$Mpc. On smaller scales, there may be non-negligible
contributions to the flow from peculiar motions generated by the
gravitational instability caused by local matter inhomogeneities.
This can lead to a non-alignment with the flow at lower $z$ in
general agreement with the trends in Fig. 1.

\acknowledgments This work is supported by NASA ADP grant
NNG04G089G and the Ministerio de Educaci\'on y Ciencia/''Junta de
Castilla y Le\'on'' in Spain (FIS2006-05319, PR2005-0359 and
SA010C05). We thank Gary Hinshaw for useful information on the
WMAP data specifics.
%\texttt{\{thebibliography\}}%

\clearpage

\end{document}